\documentstyle[multicol,aps,prb,epsfig]{revtex}

\begin{document}
 \draft

 \title{
 The Hypergraphite: A possible extension of graphitic network
 }
 \author{
 Yoshiteru Takagi, Mitsutaka Fujita\cite{memo}
 Katsunori Wakabayashi, \\
 Masatsura Igami, Susumu Okada,
 }
 \address{
 Institute of Materials Science, University of Tsukuba,
 Tsukuba 305-8573 Japan}

 \author{
 Kyoko Nakada
 }
 \address{
 College of Science and Engineering,
 Aoyama-Gakuin University, Atsugi 243-0123, Japan}

 \author{
 Koichi Kusakabe
 }
 \address{Graduate School of Science and Technology, 
 Niigata University, Ikarashi 950-2181, Japan}
 \date{\today}
\maketitle
\begin{abstract}
We propose a class of networks which can be regarded as an extension
of the graphitic network.  
These networks are constructed so that surface states with non-bonding
character (edge states) are formed in a tight-binding model with one
orbital for each atomic site.
Besides, for several networks, the tight-binding electronic
structures become a zero-gap semiconductor.  
These properties have been found in the $\pi$-electron system of the
graphene.  
Thus, we call these networks hypergraphite. 
\end{abstract}

\begin{multicols}{2}[]

\section{Introduction}

Graphite has many intriguing aspects from scientific and industrial view 
points.
It has high electronic conductivity and large anisotropic diamagnetic
susceptibility.  
The electronic structure is semi-metallic.\cite{Graphite-review,supercarbon}
An important point is that these characteristic features of an
electronic structure 
originate from a particular lattice structure of graphite.
The lattice structure is a stack of honeycomb networks of carbon atoms.
A single layer of graphite is called graphene.

Our interest in graphite has deepened much by findings of rich physics
not only in pure graphite itself but also in various graphitic materials.
For example, fullerenes\cite{Fullerene} and carbon
nanotubes\cite{Nanotube} are regarded as a deformed graphitic structure.
The discoveries have revealed that the topology of $sp^2$ carbon
networks crucially influences their $\pi$-electronic structures. 
An example showing effects of topology is that electronic structures of 
single wall carbon nanotubes depend on the chiral vectors.\cite{FSDD,HO}
The $\pi$-electronic structures show metallic or 
semi-conducting behavior. 
Interestingly, recent experiments confirmed this interplay between
the topology of $sp^2$ carbons and the $\pi$-electronic
states.\cite{stm1,stm2}

Let us start from the graphene to understand the electronic 
structures of graphite and related materials.
The electronic structure of the graphene around the Fermi energy is composed
of a $\pi$ (or $\pi^{*}$)-band.
The $\pi$-band and  the $\pi^{*}$-band are the highest valence band and 
the lowest conduction band, respectively.
The two bands degenerate at the edge of the first Brillouin
zone (1st BZ), called K (or K')-point, and show $k$-linear dispersion
at these points. 
Thus, the graphene is a zero-gap semiconductor.
Namely, the density of states (DOS) becomes zero at the Fermi energy and
linearly increases with leaving from the Fermi energy.  

Another interesting $\pi$-electron system is 
nanometer-sized graphite called ``nanographite''.\cite{Fujita}
Nanographites are a class of mesoscopic systems which are situated
between aromatic molecules and graphite.
In nanographite, the presence of edges, their shapes and size
crucially affect their $\pi$-electronic structures.

We introduced ribbon models of a graphene to study effects of the edge
on their $\pi$-electronic structures theoretically. 
The typical edge shapes are a zigzag edge or an armchair edge. 
Hence, we investigated zigzag ribbons, which are graphene ribbons with
only zigzag edge, and armchair ribbons, 
which are graphene ribbons with
only armchair edge.

There is a clear difference between $\pi$-electronic structures of these
two ribbon models.
In case of zigzag ribbons, a localized states appear around the edge at the
Fermi energy. \cite{Fujita,Tanaka,Hosoya,Kobayashi}
While such localized states do not exist on the edge of armchair ribbons.
Thus appearance of the localized states is a topological effect of
zigzag ribbons.

The localized states on zigzag ribbons were named ``edges states'', 
because they exist around edges of zigzag ribbons.
A solution of the edge states was constructed on a semi-infinite 
graphene with a zigzag edge.\cite{Fujita} 
This solution shows  that the edge states is a non-bonding orbital
(NBO).
Band structures of zigzag ribbons possess a pair of almost flat bands
near the Fermi energy.
These flat bands, {\it i.e.} edge states, induce a sharp peak in DOS at
the Fermi energy\cite{super}.  
Therefore, zigzag ribbons are expected to show unusual properties, {\it
e.g.} spin polarization,\cite{Fujita} lattice deformation,\cite{Igami} 
magnetic field effects,\cite{Wakabayashi}
transport properties,\cite{Wakabayashi2} {\it etc}.  
The appearance of edge states on zigzag ribbons is also recognized   
in terms of the
first-principles calculations within the framework of the local density 
approximation.\cite{Miyamoto,Nakada}

Now it is natural to ask a question whether other structures which
have edge states due to a topological reason exist or not.
This is because interesting phenomena have been found only for zigzag
ribbons of a graphene. 

In this paper, we report on a method to construct an extension of  the
graphitic network.
The extended networks possess the same characteristics that the graphene
shows both in the topological network and in the electronic structures. 
The networks are an AB bipartite network.
When a proper boundary condition is subjected to them, 
edge states appear in these networks. 
Besides, the bulk states of these networks become a zero-gap semiconductor.
And the highest valence band and the lowest conduction band degenerate
at the Fermi energy in 1st BZ and the two bands show $k$-linear
dispersion at the Fermi energy. 
We name the networks ``hypergraphite'', because they are regarded as
an extension of the graphitic network.

The organization of this paper is the following.
In section II, we argue about edge states.
In particular, we introduce networks in which edge states appear under a
proper boundary condition. 
In section III, we argue about bulk states of these networks by using a 
numerical calculation and an effective mass approximation.
In section IV, we give a definition of the hypergraphite.

\section{EDGE STATES}

The purpose of this section is to show several networks in which
edge states appear. 
First, we recall a solution of edge states on a semi-infinite graphene
with a zigzag edge.
Then we consider the relationship between the topology of zigzag ribbons 
and the solution of edge states.
This consideration leads us to find a method to construct these networks.
Next, we give solutions of edge states which appear in these networks.
We show band structures of slab model of them.

In this section and the next section, we use a single-band tight-binding 
model (s-TBM).
This is because we study the relation between topology of networks and 
electronic structures.
The Hamiltonian is written as
\[
H_{TBM} = -t \sum_{\langle i,j \rangle,\sigma} 
(c_{i,\sigma}^\dagger c_{j,\sigma} + {\rm h.c.} ) \; .
\]

\noindent
where $c_{i,\sigma}^{\dagger}(c_{i,\sigma})$ creates (annihilates) an
electron with spin, $\sigma$, on the $i$-th site (atom). 
We let the on-site energy zero, which is always possible for s-TBM.
A bond connection is represented by ${\langle i,j \rangle}$ and taken
only between connected sites with hopping integral, $t$.
Hereafter we assume that the hopping integral, $t$, is unity 
for simplicity.
And we consider only a paramagnetic state.
In addition, each site is occupied by one electron on the average.
  
\subsection{EDGE STATES ON A ZIGZAG RIBBON}

We recall how to construct the solution of the edge states on a
semi-infinite graphene with a zigzag edge.\cite{Fujita}
The solution, $\phi_{k,D}(x,z)$, is for the $\pi$-electron systems
(Fig.1(a)). 
Here, $k$ and $D$ denote a wave vector along the edge ($x$-direction) and
a dumping factor inward the ribbon ($z$-direction), respectively.
Each atom at the zigzag edge is assumed to be terminated by a hydrogen
atom. 
The honeycomb network is AB bipartite and each atomic site 
can be classified into A- or B-site. 
In terms of this classification, all of the zigzag-edge sites of
a semi-infinite graphene is in one of the two sublattices.
In this paper, we define sites in the sublattice containing the edge sites 
as A-sites and sites in the other sublattice as B-sites.

We list some characteristics of $\phi_{k,D}(x,z)$.
1) The amplitude is non-zero only on A-sites.
This means that edge states are NBOs.
2) Edge states are a dumping wave with a dumping factor,
$D=2\cos\frac{k}{2}$, when $\frac{2\pi}{3}<k<\pi$.
3) The edge state completely localizes at the edge, ($D=0$), when $k=\pi$.
The amplitude are +1 and -1 alternately along the edge except for
normalization.
4) The edge state approaches a propagating wave ($D\rightarrow 1$), when
$k\rightarrow \frac{2\pi}{3}$.
This limit, $\phi_{k2\rightarrow\pi /3,D \rightarrow 1}$, coincides with bulk
states at the K point in BZ of the graphene.

A construction method of the solution of the edge states in
a semi-infinite graphene with a zigzag edge\cite{Fujita}
is briefly summarized as follows.
We use a notation ($x,nj$), ($j=A$ or $B$), to specify a site.
Here, $x$ denotes a position of a cell represented by dashed lines in
$x$-direction  (See Fig. 1(b)).
The cell contains two types of sites of both A- and B-sites.
$nA$ (or $nB$) represents A sites (or B sites) in the $n$-th
zigzag-chain in the cell, respectively.
We start from setting a trial function, $\psi_{k}(x,1A)=exp(ikx)$, to
construct the solution of edge states.
Next, we determine amplitudes of $\psi_{k}(x,iA),(i=2,\cdots,\infty)$, 
so that $\psi_{k}(x,jB),(j=1,\cdots,\infty)$, become zero.
The final expression becomes an eigen function with zero amplitude on
B-sites, {\it i.e.} edge states.

\subsection{NETWORKS WITH EDGE STATES}

We consider the relationship between topology of zigzag ribbons and the 
above method to construct the wave functions.
Usually, a network of zigzag ribbons is regarded as assembly of
hexagons. 
However, we give another topological view point to the network of zigzag
ribbons. 
It is that the network of zigzag ribbons are regarded as a network made by
linking A-site 
of the $i$-th AB bipartite zigzag-chain and B-site of the ($i+1$)-st one
by extra bond ($i=1,\cdots,\infty $) (Fig. 1(c)). 
A characteristic of the view point is to connect zigzag-chains which are
an AB bipartite lattice and have a NBO as an eigen-function.   

We conjecture conditions for networks having edge states from the
topological view point as follows. 
1) A network is regarded as one made by successively connecting units.
The unit is an AB bipartite lattice which has a NBO as an eigen-function. 
2) The connection is made by linking A-sites of the $i$-th bipartite
lattice with B-sites of the ($i+1$)-st bipartite lattice by extra bonds
($i=1,2,\cdots,\infty$). 

\begin{center}
\begin{minipage}{64mm}
\epsfxsize=\hsize
\epsffile{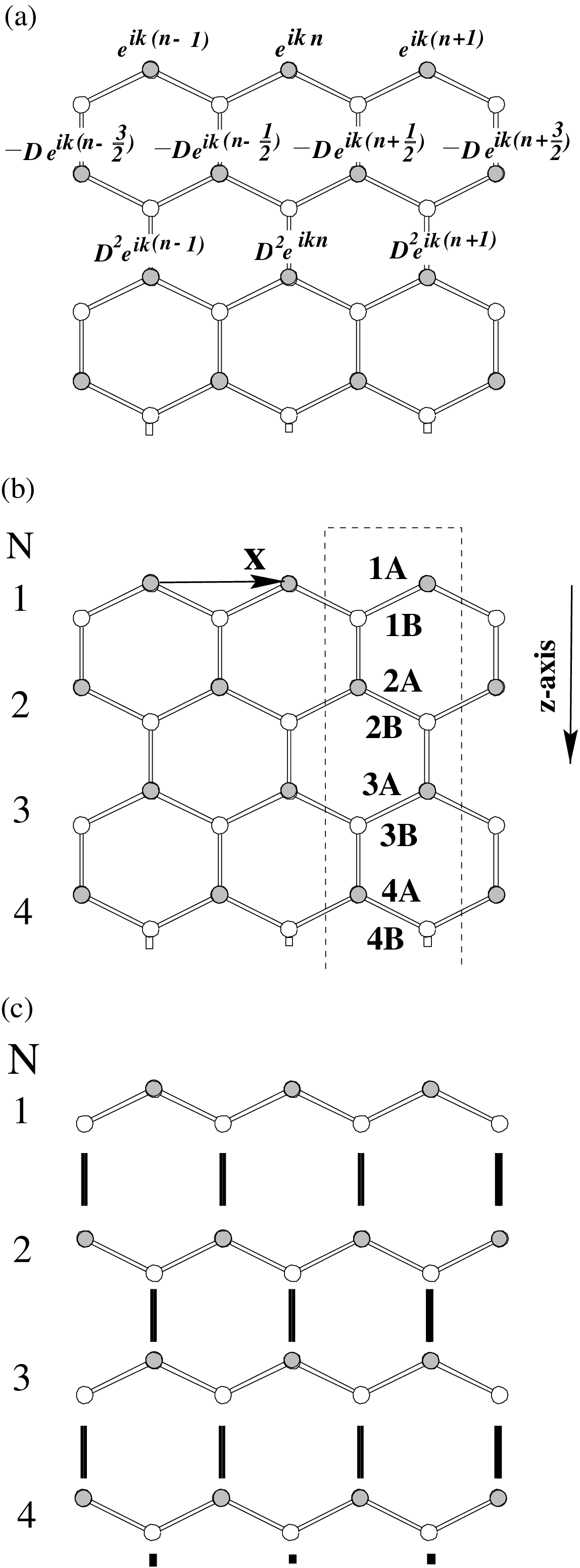} 
\end{minipage}
\end{center}
\begin{center}
\begin{minipage}{80mm}
Fig.1. (a) A solution of an edge state for a semi-infinite 
graphene with a zigzag edge.
Closed circles and opened circles represent A-sites and B-sites, respectively.
The wave function has finite amplitude at A-sites. 
$D$ is a dumping factor, $D = 2 \cos \frac{k_{x}}{2}$. 
(b) This figure shows the notation which we use in text. 
The arrow represents a unit vector. 
Dash line shows a unit cell of a semi-infinite graphene with a
zigzag edge.
$N$ represents the numbers of zigzag chains. 
(c) A way to construct a zigzag ribbon from zigzag chains. Black bonds
are extra bonds to connect zigzag chains
\end{minipage}
\end{center}

\begin{center}
\begin{minipage}{64mm}
\epsfxsize=\hsize
\epsffile{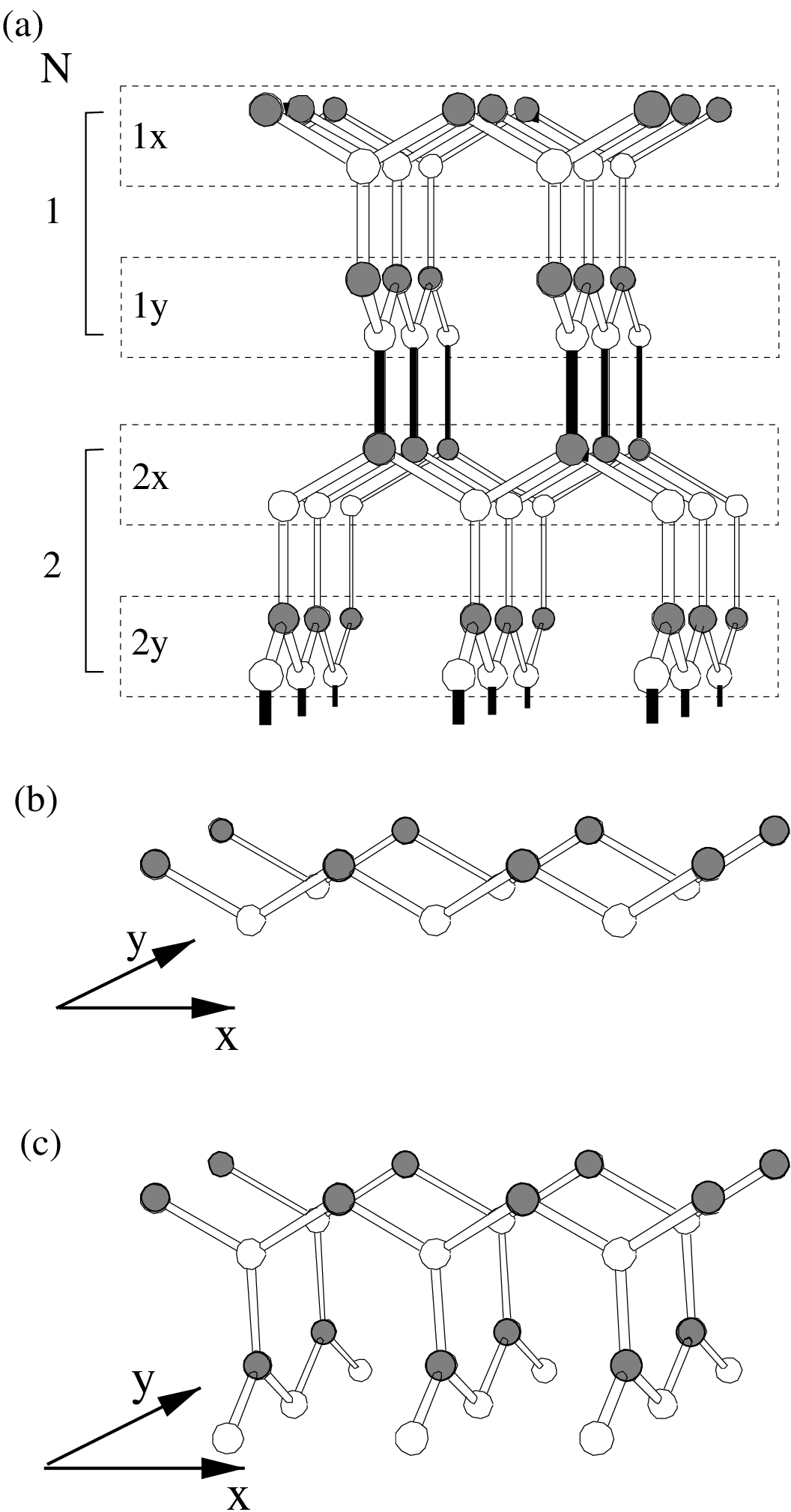} 
\end{minipage}
\begin{minipage}{80mm}
Fig.2. 
(a) The three-dimensional three-fold coordinated network (the 3-3 network).
$N$ represents the number of blocks shown in Fig. 2(c). 
Closed circles and opened circles represent A-sites and B-sites,
 respectively.
(b) 1D zigzag chains aligned in the x-direction. 
(c) A structure made by connecting zigzag chains aligned in x-direction 
and y-direction.
\end{minipage} 
\end{center}

Here, we introduce three examples which satisfy the above conditions.
They are the three-dimensional three-fold coordinated network 
(the 3-3 network), the diamond structure (DS) and the  three-dimensional 
five-fold coordinated network (the 3-5 network).
We show the 3-3 network in Fig. 2(a), the DS in Fig. 3(a), and the 3-5
network in Fig. 4.  
Of course, edge states appear in all of them under a proper boundary
condition as shown in the next subsection.

We show that they satisfy the above conditions.
The 3-3 network is regarded as a network connecting AB bipartite lattices as
shown in Fig. 2(b). 
The AB bipartite lattice is made by combining a set of zigzag chains
aligned in $x$-direction (Fig. 2(c)) and in $y$-direction, alternately.
A NBO exists on the AB bipartite lattice as an eigen function.
The DS and the 3-5 network are regarded as networks of connected honeycomb
lattices and of connected square lattices, respectively.
Naturally, a honeycomb lattice and a square lattice are an AB bipartite
lattice and have a NBO as an eigen-function.

\end{multicols}

\begin{center}
\begin{minipage}{150mm}
\epsfxsize=\hsize
\epsffile{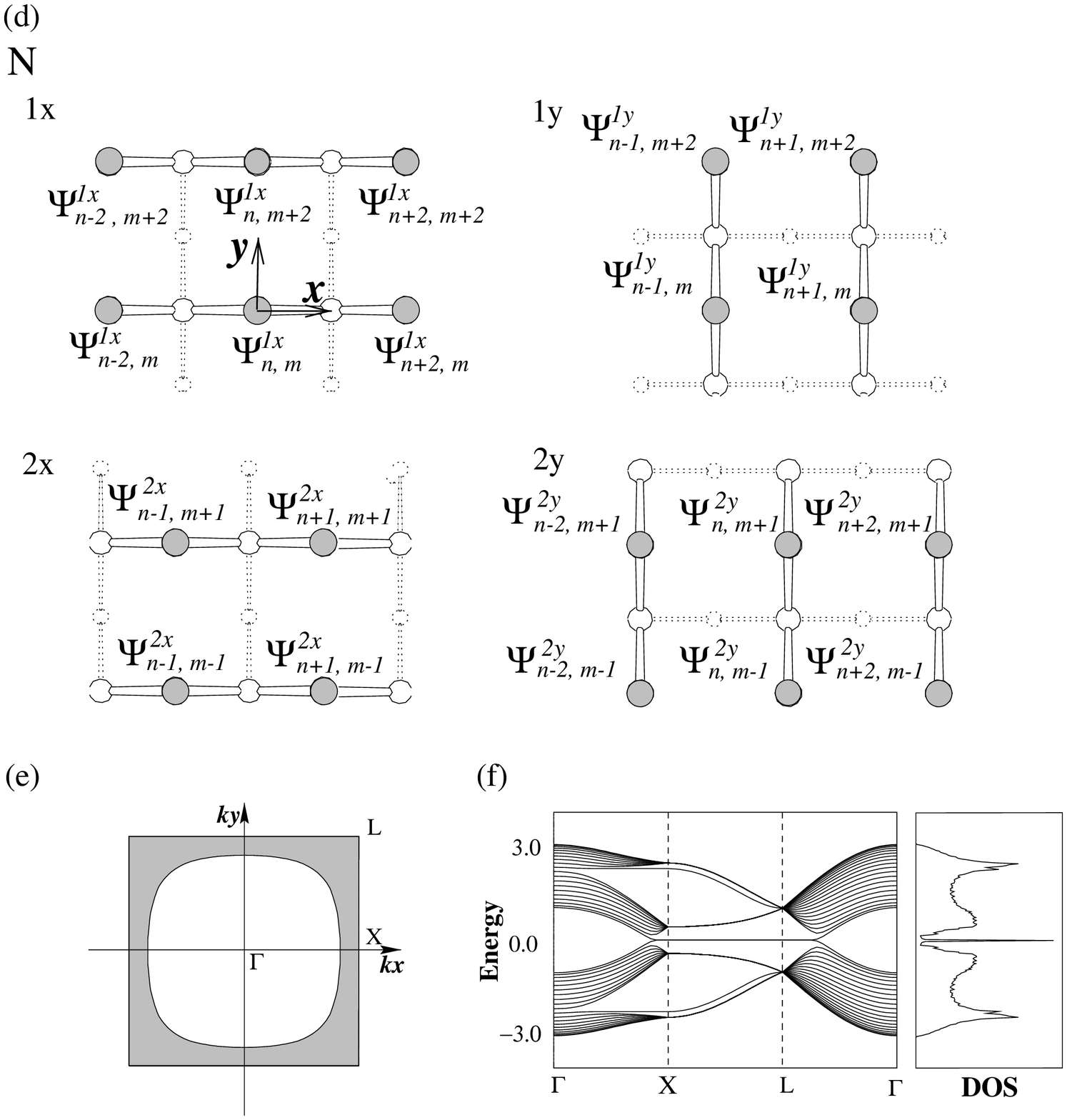} 
\end{minipage}

\begin{minipage}{160mm}
Fig.2.
(d) A solution of the edge state for the semi-infinite 
3-3 network with zigzag surface. 
The wave function of the edge states has finite amplitude
at the sites  indicated by closed circles. Dashed lines represent the
next layer  to the layer shown by solid lines.
And two arrows represent two-dimensional unit vectors.  
The wave function is given by 
 $\Psi_{n,m}^{lx}=D^{l-1}\exp(i(k_{x}n+k_{y}m))$,
 $\Psi_{n,m}^{ly}=D^{l-1}(-2\cos k_{x})\exp(i(k_{x}n+k_{y}m))$
and 
$ D= 4 \cos k_{x} \cos k_{y}$. 
$D$ is a dumping factor.
(e) The 1st BZ of a slab model of the 3-3 network. 
Edge states emerge in the shadowed region.
(f) A band structure and DOS of the 3-3 network. 
Here, we adopt a slab model with $N = 10$. 
\end{minipage}
\end{center}

\begin{multicols}{2}[]

\subsection{SOLUTIONS OF EDGE STATES}

We show solutions of edge states which appear in the 3-3 network and the DS
under a boundary condition.
The boundary condition is an open boundary condition with a surface.
The surface is composed of A-sites being only in a connected AB
bipartite lattice used as a unit to make the full network.
We call the surface ''zigzag surface''.
A dumping factor, $D$, and an area in which edge states appear in 1st BZ 
are determined from the solution.

We show solutions of edge states in Fig. 2(d) for the 3-3 network  and in
Fig. 3(b) for the DS.
In this paper, we do not show a solution of the 3-5 network, but a similar
solution is constructed for it.
Here, we note that s-TBM is used when solutions of edge states are
constructed.
Namely, we assume that a hopping integral, $t$, is the same and unity 
between any connected sites in any direction. 

\end{multicols}

\begin{center}
\begin{minipage}{150mm}
\epsfxsize=\hsize
\epsffile{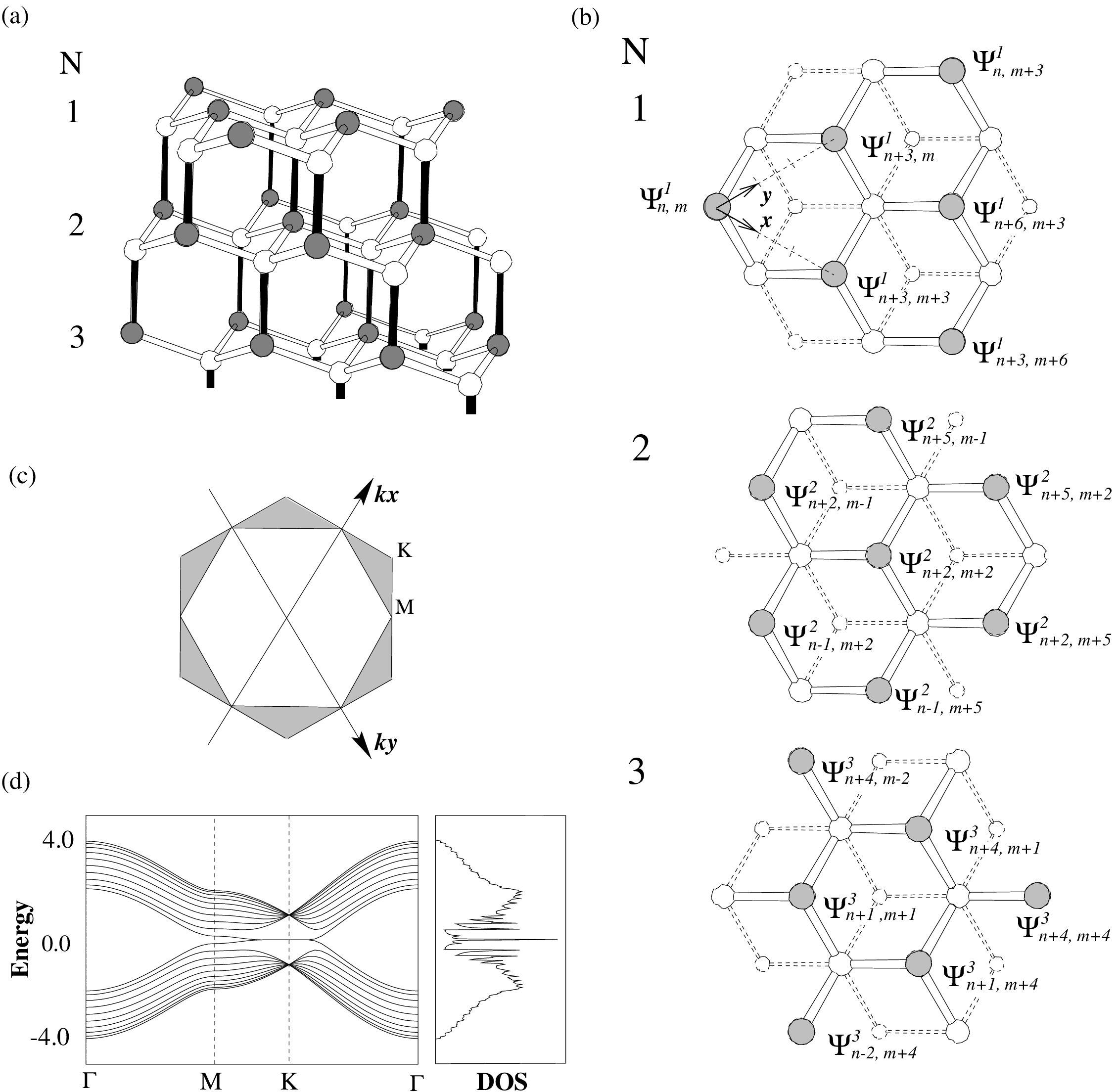} 
\end{minipage} 
\vspace*{3mm}

\begin{minipage}{160mm}
Fig.3. (a) A diamond structure. 
$N$ represents the number of honeycomb lattices. 
(b) A solution of the edge states for semi-infinite DS {\it
 i,e,} with (111)-surface. 
In each panel, dashed lines represent a next layer to a layer
represented by solid lines. 
The wave function of the edge states has the amplitude at the sites
indicated by closed circles.  
And two arrows represent two-dimensional unit vectors.  
The wave function 
is given by $\Psi_{n,m}^{l}=(-D)^{l-1}\exp(i(k_{x}n+k_{y}m))$
and $ D =   e^{ i ( - 2 k_{x} +    k_{y} ) } 
          + e^{ i (     k_{x} -  2 k_{y} ) } 
          + e^{ i (     k_{x} +    k_{y} ) }   $.
$D$ is a dumping factor.
(c) The 1st BZ of a slab model of DS. Edge states emerge 
in the shadowed region.
(d) A band structure and DOS of a slab model of DS.
Here, we adopt a slab model with $N = 20$. 
\end{minipage}
\end{center}

\begin{multicols}{2}[]

\subsection{BAND STRUCTURES OF SLAB MODELS}
The electronic band structures of a slab model are shown in Fig. 2(f) for
the 3-3 network and in Fig. 3(d) for the DS.
They are obtained by solving the eigenvalue equations of s-TBM for
each structure numerically.
Since, we consider the case that the electron number is the same as the
number of atomic sites, the Fermi energy becomes just zero for these
models\cite{bipartite}. 

A pair of almost flat bands exists at the Fermi energy in both band
structures. 
There is a sharp peak at the Fermi energy in DOS for the 3-3 network and
the DS. 
We recognize appearance of edge states from these results. 

Edge states appear in the 3-3 network and the DS which satisfy the conditions
described in subsection B.
In this paper, we show only two examples, but we have found many
networks which satisfy the conditions and confirmed that edge states
appear in these networks.
In addition, the dimension of the structure does not have to be three or
two. 
It is possible to design a higher dimensional structure with edge states.
A dumping factor for these networks is obtained by using a transfer
matrix method (See Appendix. A).
 
\begin{center}
\begin{minipage}{64mm}
 \epsfxsize=\hsize
 \epsffile{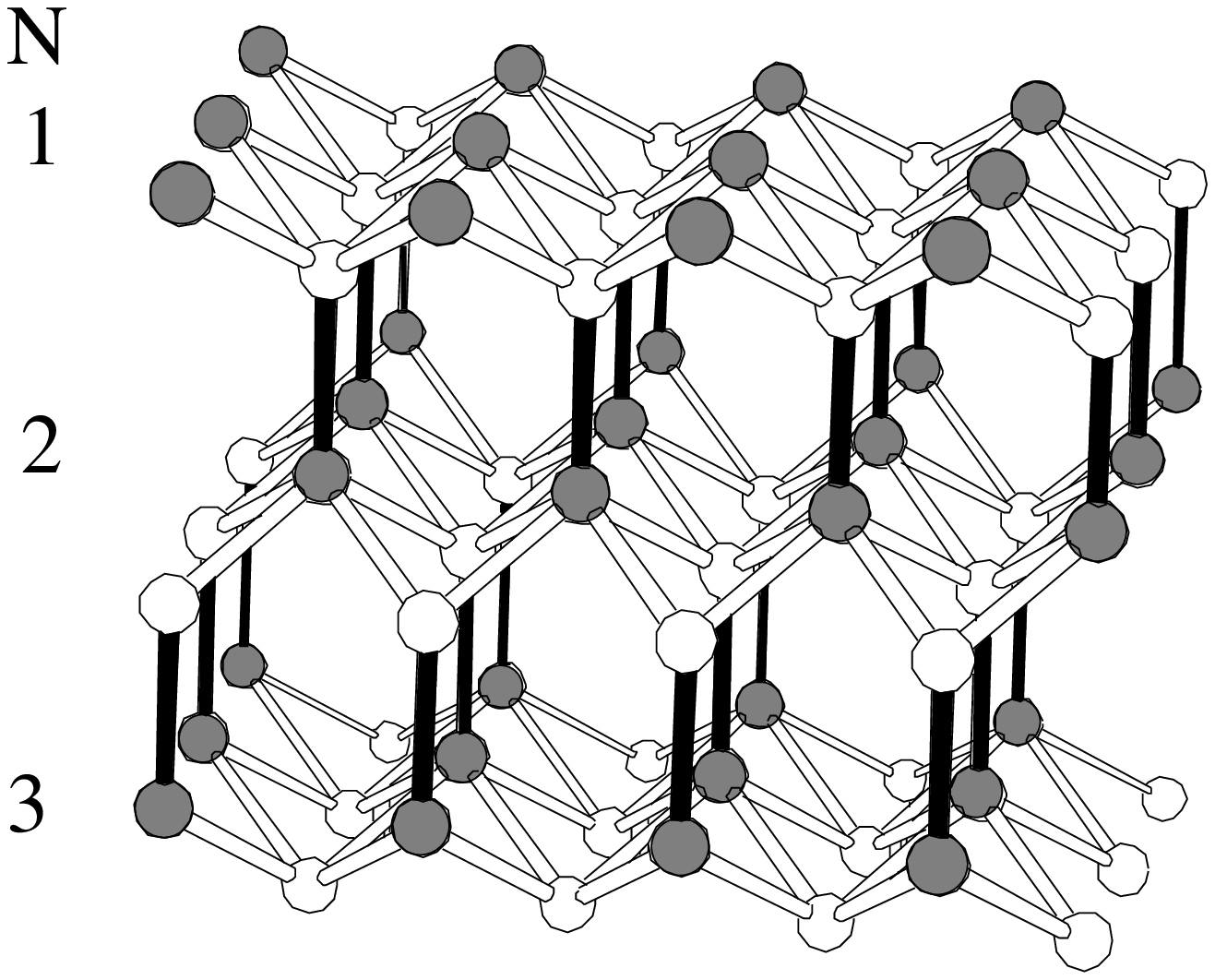}
\end{minipage}
\begin{minipage}{80mm}
Fig.4. A three-dimensional five-fold coordinated network.
$N$ represents the number of square lattices. 
\end{minipage}  
\end{center}

\section{BULK STATES}

In this section, we argue about bulk states of the networks which satisfy
the conditions described in the last section.
Electronic structures of the 3-3 network and the DS with a zigzag surface
are similar to that of the graphene. 
Here, we have a question of how the electronic structure 
of bulk states is.
Bulk states of the 3-3 network and the DS are made use of in this section as
examples.  

\subsection{BAND STRUCTURES}

We show electronic band structures and DOS in Fig. 5(a) for the 3-3
network and in Fig. 6(a) for the DS, respectively.
The Fermi energy is just zero in the model. 

At first, we explain the band structure of the 3-3 network.
As shown in Fig. 5(a), the lowest conduction band and the highest valence
band degenerate at a point on the $\Gamma -X$ line.
At this point, two bands have $k$-linear dispersion.
This is the same characteristic that the $\pi$-states of the
graphene have.
But, in the case of the 3-3 network, the lowest conduction band and the 
highest valence band degenerate on a closed line in 1st BZ.
This line is shown in Fig 5(b).
Recall that the lowest conduction band and the highest valence band
degenerate at only $K$ (or $K'$) point in the case of the graphene.
This line for the 3-3 network is understood to be an extension of the
$K$ point of the graphene.
Hence, we call this line ``$k$-line''.
This point is argued by using the $k\cdot p$ approximation in next
subsection again.
In addition, the system is a zero-gap semiconductor as seen in DOS.

Next, we explain the band structure of the DS.
There is a $k$-line, where the lowest conduction band and the highest
valence band degenerate and two bands possess $k$-linear dispersion at
W point. 
These $k$-lines are shown in Fig. 6(b).
The system is also a zero-gap semiconductor as seen in DOS.

\begin{center}
\begin{minipage}{75mm}
\epsfxsize=\hsize
\epsffile{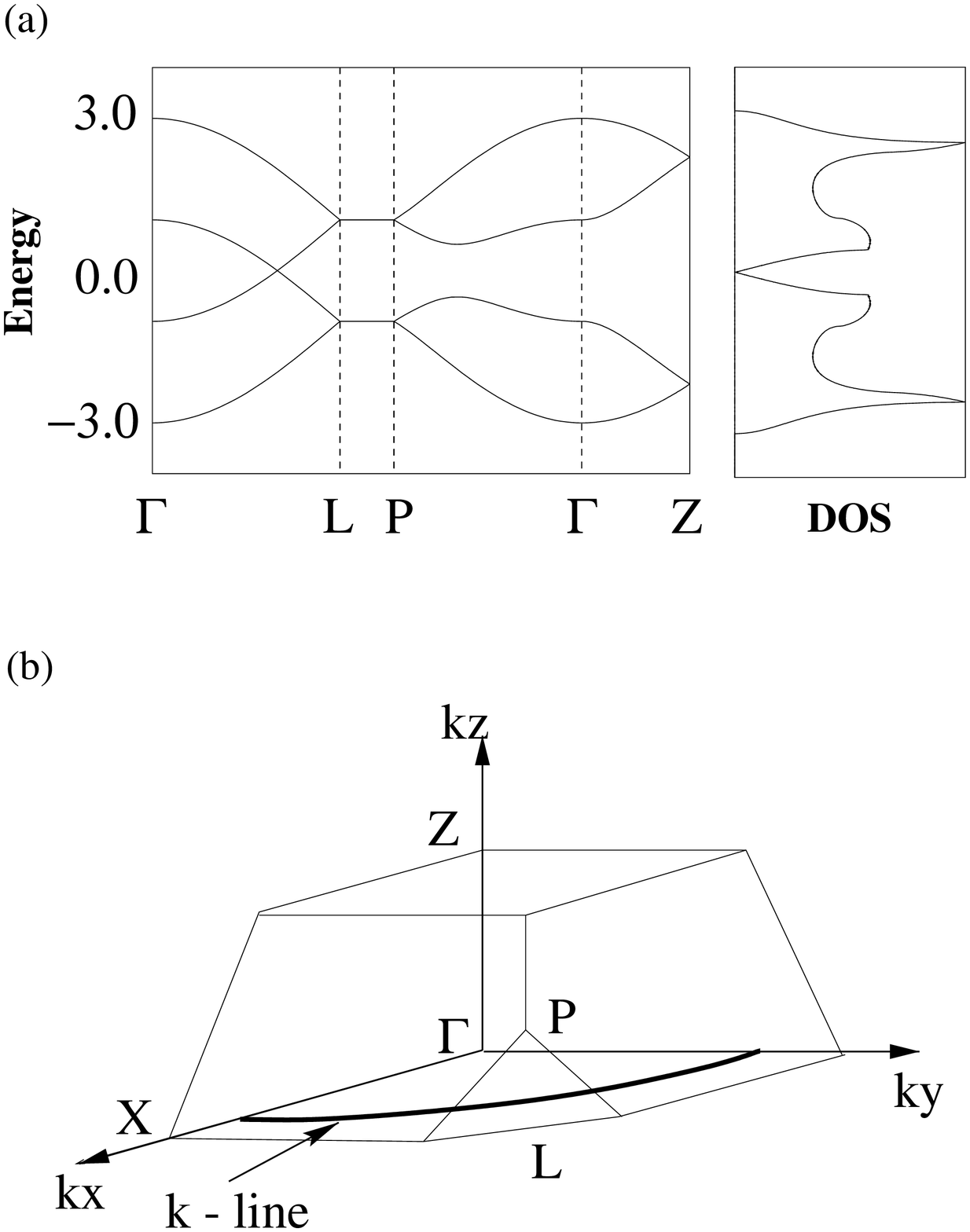}
\end{minipage}

\begin{minipage}{80mm}
Fig.5. (a) A band structure and DOS of a bulk of the 3-3 network. 
(b) The 1st BZ of the 3-3 network.  
In this figure, a region which contain 
$k_{x} \geq 0 $ and $ k_{y} \geq 0 $ is showed. 
The $k$-line is represented by the thick line.
\end{minipage}  
\end{center}

\begin{center}
\begin{minipage}{75mm}
\epsfxsize=\hsize
\epsffile{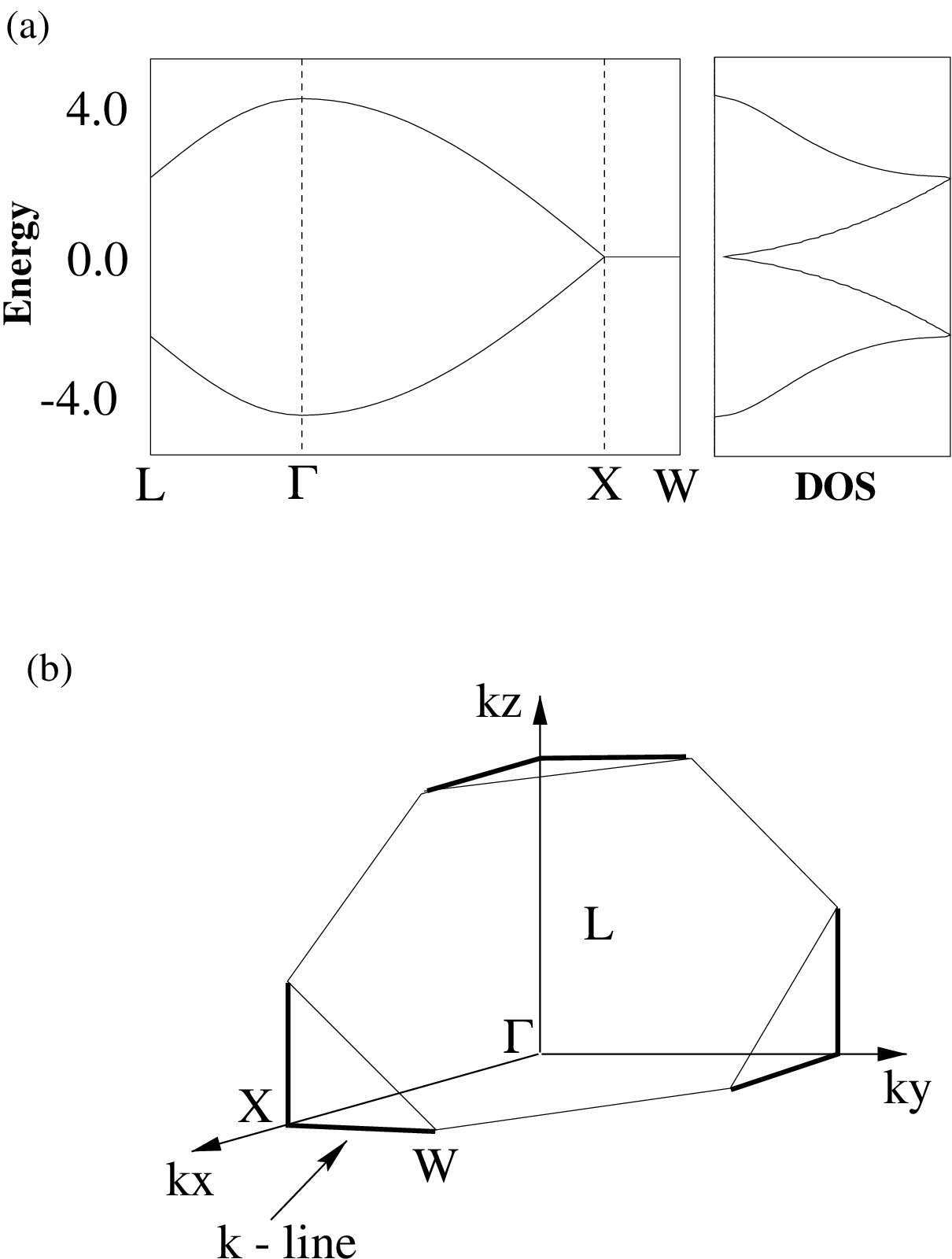}
\end{minipage}
\begin{minipage}{80mm}
Fig.6. (a) A band structure and DOS of a bulk of the DS. 
(b) The 1st BZ of the DS. 
In this figure, a region which contain 
$k_{x} \geq 0 $ and $ k_{y} \geq 0 $ is showed. 
The $k$-lines are represented by the thick lines.
\end{minipage}
\end{center}

The electronic structures of both the 3-3 network and the DS show the same
characteristic. 
However, there is a difference in the shape of $k$-line of these two
structures. 
The $k$-line is closed in 1st BZ in a case of the 3-3 network, while the
$k$-line is open in 1st BZ in a case of the DS.

We have studied many networks which satisfy the conditions 
described in the last section except the 3-3 network and the DS. 
The most of the networks show the same characteristics that 
the 3-3 network or the DS shows.
One point which we have to remark is that in some structures made by our 
construction method, the bulk band structure is not a zero-gap
semiconductor but a metal.
In these exceptional cases, edge states appear in these networks whose
bulk states become metallic.

\end{multicols}

\subsection{K$\cdot$P APPROXIMATION}

In this subsection, we study bulk states of the 3-3 network and the DS
using the $ k \cdot p$ approximation \cite{Kohn}. 
The results show again that there are bands with $k$-linear dispersion in
these networks.

First, we study the 3-3 network by the $k \cdot p$ approximation. 
We start from obtaining wave functions on a $k$-line by diagonalizing 
the single-band tight-binding Hamiltonian. 
Wave functions are represented in a Bloch form as 
\begin{eqnarray} 
\psi_{{\bf k}}({\bf r}) = \frac{1}{\sqrt{N}} 
\sum_{j=1}^4 \sum_{{\bf R}_{i}} 
\exp({i {\bf k} \cdot {\bf R}_{i}})b_{j}({\bf k})  
\phi_{j}({\bf r-R}_{i}), 
\end{eqnarray}
where $\phi_{j}({\bf r-R}_{i})$ is a wave function of the $j$-th
site in a unit cell.  
The vector, $ {\bf R}_{i} $,  represents
the position of the $i$-th cell. 
$ b_{j}({\bf k}) $ is an amplitude which should be determined. 
$N$ is a number of unit cells.
Here, we assume that $\phi_{j}({\bf r-R}_{i})$
are orthonormal: 
\[
 (\phi_{j}({\bf r}-{\bf R}_{i}),
  \phi_{j'}({\bf r}-{\bf R}_{i'})) = 
\delta_{{\bf R}_{i},{\bf R}_{i'}}
\delta_{j,j'}
.
\]

The eigenvalue equation of the 3-3 network becomes following, 
\begin{eqnarray}
H({\bf k}) {\bf b} ({\bf k}) = \varepsilon {\bf b} ({\bf k}). 
\label{eigen}
\end{eqnarray}
The matrix, $H({\bf k})$, is defined by, 
\begin{eqnarray} 
 H({\bf k}) = \left(  \begin{array} {cc}
          {\bf 0 }     & {\bf A } \\ 
          {\bf A^{\dagger} } & {\bf 0 } 
             \end{array}
     \right),
\label{3H1}
\end{eqnarray}
where 
\begin{eqnarray}
 {\bf A } = 
   \left( 
    \begin{array}{cc}
       - (1+ \exp (i \sqrt{3}k_{x}a) &
    - \exp (i(\frac{\sqrt{3}}{2}k_{x}a+\frac{\sqrt{3}}{2}k_{y}a+ 3 k_{z}a) \\
       -1    &
       - (1+ \exp (i \sqrt{3}k_{y}a))
    \end{array} 
   \right).
\end{eqnarray}
{\it a} is the bond length. 
A vector ${\bf b}({\bf k})$ is given by
\[
   {\bf b} ({\bf k}) = \left( \begin{array}{c}
                      b_{1}({\bf k}) \\
                      b_{2}({\bf k}) \\
                      b_{3}({\bf k}) \\
                      b_{4}({\bf k})
                           \end{array}
                    \right).
\]

We can deduce equations determining a $k$-line 
on which energy is equal to zero from the above eigenvalue equation (Eq.
(\ref{eigen})).  
The condition, ${\rm det} {\bf A} =0$, 
gives $k$-line equations, 
\begin{eqnarray}
  \left\{ 
     \begin{array}{l}
          \cos \frac{\sqrt{3}}{2}K_{x}a  \cos \frac{\sqrt{3}}{2}K_{y}a 
                                                         = \frac{1}{4} \\
          k_{z}=0
      \end{array}
   \right.,
\label{3line}
\end{eqnarray}  
where ${\bf K} =(K_{x},K_{y},0)$ is a point on $k$-line.

We need wave functions on $k$-line to apply the  $k\cdot p$ approximation. 
These are given in a Bloch form as, 
\begin{eqnarray}
 \psi_{A}({\bf r}) &=& \frac{1}{L\sqrt{N}}
  \sum_{{\bf R}_{i}} 
  \{ \exp ( i {\bf K} \cdot {\bf R}_{i}) 
     (- \exp(-i \beta))Y \phi_{1}({\bf r-R}_{i}) 
      +             X \phi_{2}({\bf r-R}_{i})) \},
     \nonumber \\
 \psi_{B}({\bf r}) &=& \frac{1}{L\sqrt{N}}
  \sum_{{\bf R}_{i}} 
  \{ \exp ( i {\bf K} \cdot {\bf R}_{i})
     (             Y \phi_{3}({\bf r-R}_{i}) 
      - \exp( i \alpha)X \phi_{4}({\bf r-R}_{i})) \} 
    \; , \nonumber
\end{eqnarray}
where 
\[
X=\sqrt{2 \cos \alpha}, ~~~~ 
Y=\sqrt{2 \cos \beta }, ~~~~ 
L=\sqrt{2 \cos \alpha+2 \cos \beta},
\]
and 
\[
\alpha =\frac{\sqrt{3}}{2}K_{x}a, ~~~~ 
\beta  =\frac{\sqrt{3}}{2}K_{y}a. 
\]

We introduce a wave vector {\bf k} measured from a point on $k$-line 
and define two functions, 
\begin{eqnarray}
  \Phi_{j,{\bf k}}({\bf r})=\exp(i {\bf k \cdot r}) \psi_{j}({\bf r}),
\end{eqnarray}
with $j=A,B$. These functions are orthonormal:
\[
 (\Phi_{j,{\bf k}},\Phi_{j',{\bf k'}})=\delta_{jj'}\delta_{{\bf k}{\bf k'}}.
\] 
The wave function near the point can be expanded as 
\begin{eqnarray}
  \Psi({\bf r})=\sum_{j=A,B} \int \frac{d{\bf k}}{(2 \pi )^{3}} 
                 C_{j}({\bf k}) \Phi_{j,{\bf k}}({\bf r})
               =\sum_{j=A,B} F_{j}({\bf r}) ~ \psi_{j}({\bf r}),
\label{extend}
\end{eqnarray}
where $F_{j}({\bf r})$ is the envelop function defined by 
\begin{eqnarray}  
   F_{j}({\bf r})=\int \frac{d{\bf k}}{(2 \pi )^{3}} 
                  \exp(i {\bf k \cdot r})  C_{j}({\bf k}).
\label{Envelop}
\end{eqnarray}  
Substituting Eq.(\ref{extend}) into the Schr\"{o}dinger equation, we 
obtain the following $ k \cdot p $ equation,
\begin{eqnarray}
\left\{ 
   \begin{array}{l} 
 \frac{1}{L^2} \exp(i(\alpha+\beta))
  (-\frac{\sqrt{3}}{2}a \tan \alpha ~\hat{k_{x}}  
   -\frac{\sqrt{3}}{2}a \tan \beta  ~\hat{k_{y}}  
   +i~ 3a  ~\hat{k_{z}} )F_{A}({\bf r}) = E F_{B}({\bf r}) 
\\
 \frac{1}{L^2}\exp(-i(\alpha+\beta))
( -\frac{\sqrt{3}}{2}a \tan \alpha  ~\hat{k_{x}}  
  -\frac{\sqrt{3}}{2}a \tan \beta ~\hat{k_{y}}  
  -i~ 3a  ~\hat{k_{z}} )F_{B}({\bf r}) = E F_{A}({\bf r}) 
   \end{array}
\right. ,
\label{3kp1}
\end{eqnarray}
where $\hat{k}_{x,y,z}$ are defined as 
$\hat{k}_{x,y,z}=-i\nabla_{x,y,z}$. 
Here, we rewrite the above expression with 
$\hat{k_{\bot}}$ and $\hat{k_{\|}}$, 
which are components of the operator, $\hat{{\bf k}}$, 
perpendicular to $k$-line and parallel to $k$-line in $k_{x}k_{y}$-plane, 
respectively. They are given by 
\[ \left ( \begin{array}{rr} \hat{k}_\bot \\ \hat{k}_\| \end{array} \right )
   = \left ( 
         \begin{array}{rr}
          \cos \theta_{{\bf K}} & \sin \theta_{{\bf K}} \\
         -\sin \theta_{{\bf K}} & \cos \theta_{{\bf K}} 
         \end{array}
        \right )
   \left ( \begin{array}{rr} \hat{k}_x \\ \hat{k}_y \end{array} \right ),
   ~~~~~~~~
   \left \{ 
         \begin{array}{c}
           \cos \theta_{{\bf K}} = \frac{1}{ \Delta_{{\bf K}} } 
                               \tan \alpha  \\
           \sin \theta_{{\bf K}} = \frac{1}{ \Delta_{{\bf K}} } 
                               \tan \beta 
          \end{array}
   \right.,
\]
\[
   \Delta_{{\bf K}} = \sqrt{ \tan^2 \alpha +
                            \tan^2 \beta  }.
\]
By using the components, $\hat{k}_\bot$ and $\hat{k}_\|$, 
Eq. (\ref{3kp1}) becomes 
\begin{eqnarray}
 \left ( 
     \begin{array}{cc}
       0 & \exp(i(\alpha+\beta))
           (\xi_{{\bf K}} \hat{k_{\bot}} - i \eta \hat{k_{z}})   \\
      \exp(-i(\alpha+\beta))
           (\xi_{{\bf K}} \hat{k_{\bot}} + i \eta \hat{k_{z}})   & 0 
      \end{array}
    \right)
    \left(
       \begin{array}{c}
         F_{A}({\bf r}) \\
         F_{B}({\bf r})
        \end{array}
    \right)  
      =  E
     \left(
       \begin{array}{c}
         F_{A}({\bf r}) \\
         F_{B}({\bf r})
        \end{array}
    \right),
\label{3kp2}  
\end{eqnarray}
where 
\[  \left \{ 
      \begin{array}{l}
     \xi_{{\bf K}} = - \frac{\sqrt{3}}{2} a \Delta_{{\bf K}} / L^2 \\
     \eta      = 3a / L^2 
       \end{array}
     \right. .
\]
The final expression is the same as that of the $k \cdot p$ equation 
for the graphene around K (or K') point. 
We can obtain the energy eigenvalue around $k$-line from (\ref{3kp2}), 
which is  
\begin{eqnarray}
E=
\sqrt{\xi_{{\bf K}}^2 k_{\bot}^2+\eta^2 k_{z}^2}.
\label{3eg}
\end{eqnarray}
Here, $k_{\bot}$ and $k_{z}$ are a component of the wave vector, 
${\bf k}$, defined above.  
Thus, the energy dispersion around a $k$-line is in proportion to $k$, 
which is a distance from the $k$-point to the $k$-line in plane 
perpendicular to k-line.

Next, we study the DS by using the $k \cdot p$ approximation. 
We have several $k$-lines in 1st BZ in this case.
We choose a $k$-line,
\begin{eqnarray}
\left\{ \begin{array} {l}
     K_{y} = 0 \\ 
     K_{z} = \frac{ 2 \pi}{a} 
        \end{array} 
\right. ,
\label{Dkline}
\end{eqnarray}
to demonstrate the $k \cdot p$ approximation.
From the same procedure, we can obtain the following $k \cdot p $
equation, 
\begin{eqnarray}
  \left( \begin{array}{cc}
      0     &   \exp (i\frac{K_{x}a}{4} )
         (\xi_{K_{x}} \hat k_{y} + i \eta_{K_{x}}\hat k_{z})     \\
   \exp (-i\frac{K_{x}a}{4} )
     (\xi_{K_{x}}  \hat k_{y} - i  \eta_{K_{x}}\hat k_{z})  &    0   
           \end{array}
    \right)
    \left( \begin{array}{c}
             F_{A}({\bf r})     \\
             F_{B}({\bf r})   
           \end{array}
    \right)   =
    E     \left( \begin{array}{c}
             F_{A}({\bf r})     \\
             F_{B}({\bf r})   
           \end{array}
    \right),    
\label{Dkp2}
\end{eqnarray} 
where 
\[ \xi _{K_{x}} =  a \sin ( \frac{K_{x}a}{4}),~~~
   \eta_{K_{x}} =  a \cos ( \frac{K_{x}a}{4}).
\] 

This equation is again the same form as that of the graphene.
We can obtain the energy eigenvalue around the $k$-line from Eq.(\ref{Dkp2}),
which becomes 
\begin{eqnarray}
        E =\sqrt{ \xi_{K_{x}}^2 k_{y}^2 + \eta_{K_{x}}^2  k_{z}^2}.
\end{eqnarray}
Since we consider k-line which is parallel to $k_{x}$-axis, this energy
eigenvalue is proportional to a component $k$ of 
the wave number, ${\bf k}$, in a plane perpendicular to k-line,
{\it i.e} the $k_{y}k_{z}$-plane.
The point ($0,0,\frac{2\pi}{a}$), where Eq.(\ref{Dkp2}) possesses a
singularity, is an exception.
The energy eigenvalue is proportional to only $k_{z}$ on this point,  
because $\alpha_{K_{x}}$ becomes zero.  

Dispersion on a $k$-line is proportional to 
a wave number for all direction
in plane being perpendicular to the $k$-line except for some singular
points. 
From this results, $k$-line is regarded as locus of a point where
conduction and valence bands degenerate with $k$-linear dispersion, 
{\it i.e.} the $K$ (or $K'$) point of the BZ in the graphene. 
This is a reason why we use a word `` $k$-line''.

\begin{multicols}{2}[]

\section{DISCUSSION}

In this paper, we use a single-band tight binding model for explaining
nature of networks which satisfy the conditions described in the
section II{\bf B}.
In the model, we assume that the hopping integral, $t$, is the same in
any direction, in order to make a simple argument.
But, this assumption is not necessarily required for networks having edge 
states.
In a selected class of networks, we can prove the existence of
edge states on them (See appendix A).

Here, hypergraphite is defined to be a network which satisfies following 
two conditions.
One is on the topology of a network given by hopping integrals and the
other is on the electronic structure.

The topology of network given by the hopping integrals is regarded as
being constructed by the following method.  
\renewcommand{\labelenumi}{\arabic{enumi})}
\begin{enumerate}
\item Prepare an ($N-1$)-dimensional AB bipartite network having a NBO
      whose amplitude is finite only on A-sites.  
      The number of A-sites and that of B-sites in a unit cell are equal. 
\item Line up the copies of the network along z-direction. 
      We define A-sites and B-sites on the $i$-th AB bipartite
      electronic network as the same way on the first network.  
\item Connect B-sites on the first network with A-sites on 
      the second network by extra bonds. 
\item Connect B-sites on the $i$-th network with A-sites on the
      ($i+1$)-st network by the same way which is used in 3) 
      ($i=1,2,\cdots,\infty$). 
\end{enumerate}

Next, consider the s-TBM in the network, the electronic structure  
constructed above fashion has to show following characteristics.

\begin{enumerate}
\item Edge states appear in this network with a zigzag surface. 
\item The highest valence band and the lowest conduction band degenerate 
      at the Fermi energy and the Fermi surface becomes an
      ($N-2$)-dimensional surface.  
\item The highest valence band and the lowest conduction band possess
      $k$-liner dispersion at  points on the ($N-2$)-dimensional Fermi
      surface. 
\item The bulk system is a zero-gap semiconductor. 
\end{enumerate}
Here, note that the Fermi level is set on the center of the band
structure, because we consider the case of a half-filled AB bipartite
network. 

From the above definition on a network, the appearance of
edge states on hypergraphite itself is natural.  
This is because a NBO which exists on a basic AB bipartite network
coincides with a completely localized edge states in a hypergraphite.  
But it is not clear whether hypergraphite become a zero-gap semiconductor
and whether it has bands with $k$-linear dispersion at the Fermi energy.
This important point is left to be solved in general.  

Here, we list possible scenarios how hypergraphite networks and their
characters are realized in real materials.

We have shown that the three-dimensional three-fold coordinated network 
and the diamond structure are classified as the hypergraphite. 
From the point of geometric topology, the cubic diamond and the
cubic silicon possess the topology of the diamond structure. 
We also find that the 3-3 network is created by silicon in
$\alpha$-ThSi$_{2}$.

In these real materials, the electronic and geometric structures are
determined by not only $s$-electrons but also $p$-electrons of carbon or
silicon atoms. 
Actually, $sp^{3}$ or $sp^{2}$ hybridized orbitals form these
interesting lattice structure and electronic properties.
Hence, no electronic characteristic of hypergraphite is found
around the Fermi energy.
But, we notice the characteristics of the hypergraphite in the lowest
two bands.
Examples are 1) gap closing at X-point in the lowest two
$sp^3$-hybridized bands of diamond or silicon\cite{diamond}
and 2) gap closing at a point on the $\Gamma$-X line in the lowest two
$sp^2$-hybridized bands of CaSi$_2$ in the $\alpha$-ThSi$_2$
structure\cite{Kusakabe}. 
This is because the two bands are mainly composed of 2$s$ or 3$s$
electrons.

\section{SUMMARY}

We propose a class of $N$-dimensional networks.
These networks have the same characteristics of topology of hopping
integrals and an electronic structure that the graphene has.
A characteristic of topology of hopping integrals is AB bipartite network.
Particular properties of the electronic structure are 1) appearance of
edge states under a zigzag surface and 2) existence of a
$k$-line with $k$-linear dispersion, and 3) being a zero-gap semiconductor.
From the above reason, we regard the class of networks as an extended
graphitic network and name them ``hypergraphite''.

\section{ACKNOWLEDGMENTS}
The author would like to thank M. Fujita, M. Igami, K. Wakabayashi, S. Okada,
K. Nakada and K. Kusakabe.
Numerical calculations were performed on the Fujitsu VPP500 of computer
center at Institute for Solid States Physics, University of Tokyo and 
the NEC SX3 of computer center at Institute for Molecular Science, Okazaki 
National Institute. 
This work is supported by Grant-in-Aid for Scientific Research
nos.10309003 and 11740392
from the Ministry of Education, Science and Culture, Japan.

\end{multicols}

\appendix
\section{Transfer Matrix Method}

In this appendix, we show that the edge states appear in a selected
class of $N$-dimensional networks which are composed of
($N-1$)-dimensional AB bipartite network.
Naturally, we consider the $N$-dimensional networks which satisfy the
conditions of hypergraphite.

We consider a semi-infinite slab model, on which the single-band tight-
binding model is constructed.  
In this model, hopping integrals do not have to be unity.
We call an $i$-th ($N-1$)-dimensional network
$\Lambda^{i}$. ($i = 1,2,\cdots,n,\cdots,\infty$) 
It is assumed that there exists a NBO on $\Lambda^1$. 
We consider only AB bipartite networks, where 1) the number of A-sites
and that of B-sites are the same and 2) the NBO has a finite amplitude
on every A-site of $\Lambda^1$ and all of B-sites are node for the NBO. 
Since all of $\Lambda^i$ has the same network as that of $\Lambda^1$, 
all of $\Lambda^i$ has a NBO which is the copy of the first NBO. 
Then, we naturally define A-sites and B-sites on $\Lambda^i$ so that
B-sites  are always node for the NBO. 
A surface of the network is composed by only A-sites on $\Lambda^{1}$. 

We also assume that the ($n-1$)-dimensional AB bipartite network is a 
lattice having a unit cell and is periodic in all ($n-1$)-directions. 
The numbers of A-sites and B-sites in the unit cell of $\Lambda^{i}$
are equal and represented by $N_0$. 
Then, we can obtain wave functions on $\Lambda^{i}$ in a Bloch form as,
\begin{eqnarray}
   |\psi_i ({\bf k}) \rangle 
   = \sum_{j,l} \sum_{{\bf r}_i} u_{i,j,l} ({\bf k}) 
       \exp (i {\bf k} \cdot {\bf r}_i)
       c^\dagger_{i,{\bf r}_i,j,l} |0\rangle .
\end{eqnarray}
Here ${\bf k}$ and ${\bf r}_i$ represent an ($N-1$)-dimensional wave
vector and a position vector for a unit cell on $\Lambda^1$,
respectively. 
Two labels, $j$ and $l$, indicate sublattices ($j=$ A or B) and a site
of the $j$-sublattice in the cell ($l=1,\cdots, N_0$), respectively. 
An operator, $c^\dagger_{i,{\bf r}_i,j,l}$, creates an electron on a
site  indexed by $j$ and $l$ in a cell at ${\bf r}_i$ on $\Lambda^i$.

We introduce an $N_0$-dimensional vector 
${\bf u}_{i,j} = (u_{i,j,l}({\bf k}))$. 
Then we can rewrite the Schr\"{o}dinger equation on the total
$N$-dimensional  network in a form as, 
\begin{eqnarray}
\begin{array}{ccl}
\epsilon ~{\bf u}_{1,A} &=&  T^{in} {\bf u}_{1,B} , \\
\epsilon ~{\bf u}_{i,A} &=&  T^{in} {\bf u}_{i,B}+
                             T^{ex} {\bf u}_{i-1,B} , \\
\epsilon ~{\bf u}_{i,B} &=&  T^{in\dagger} {\bf u}_{i,A}+
                             T^{ex\dagger} {\bf u}_{i+1,A}  \; .
\end{array}
\label{s1}
\end{eqnarray}
Here, matrices, $T^{in} $ and $T^{ex}$, are functions of {\bf k}. 
An ($ N_0 \times N_0 $) matrix $T^{in}$ represents the transfer from
B-sites  to A-sites in $\Lambda^{i}$. 
$T^{ex}$ is another ($N_0 \times N_0$) matrix representing  transfers 
from B-sites in $\Lambda^{i-1}$ to A-sites in $\Lambda^{i}$.
$T^{in\dagger}$ is a matrix representing transfers from A-sites to
B-sites  in $\Lambda^{i}$.

In the first place, we discuss the simplest case, {\it i.e.} the unit
cell in $\Lambda^i$ has only one A site and one B site. 
Then $T^{in}, T^{ex}$ and ${\bf u}_{i,j}$ become a scalar. 
($i = 1,2,\cdots,n,\cdots \infty,~j=A ~ or ~ B$).   
We seek for a dumping wave with $\epsilon = 0$. 
Then we have the following equations with a transfer matrix, $T_{j}$,
\begin{eqnarray}
\left( \begin{array}{c}  
u_{(i+1),A} \\ u_{i,B} \end{array} \right)
=
T_{A}\left( \begin{array}{c} 
u_{i,B} \\ u_{i,A} \end{array} \right)
=
\left( \begin{array}{cc}
          0          &     - \frac{T^{in}}{T^{ex}}  \\
          1          &       0                                 
        \end{array} \right)
\left( \begin{array}{c}  u_{i,B} \\ u_{i,A} \end{array} \right) , 
\label{t1}
\end{eqnarray}
\begin{eqnarray}
\left( \begin{array}{c}  u_{i,B} \\ u_{i,A} \end{array} \right)
=
T_{B}\left( \begin{array}{c}  u_{i,A} \\ u_{i-1,B} \end{array} \right)
=
\left( \begin{array}{cc}
          0          &     - \frac{T^{ex *}}{T^{in *}}  \\
          1          &       0                                 
        \end{array} \right)
\left( \begin{array}{c}  u_{i,A} \\ u_{i-1,B} \end{array} \right), 
\label{t2}
\end{eqnarray}
with another condition, $u_{1,B}=0$. 
Here, we assume that $T^{ex}$ and $T^{in}$ are not equal to zero. 
This is natural because, a) if $T^{in}$ is zero, there exists a
completely localized edge state for a given ${\bf k}$, and b) if
$T^{ex}=0$, there is no solution with $u_{1,A}\neq 0$ and $u_{1,B}=0$. 
The problem is that on what conditions this set of equations has a dumping 
wave toward the $z$-direction. 
If we multiply two transfer matrices, we have,
\begin{eqnarray}
\left( \begin{array}{c}  u_{i+1,B} \\ u_{i+1,A} \end{array} \right)
=
\left( \begin{array}{cc}
         - \frac{T^{ex*}}{T^{in*}} &  0                         \\
                  0              & - \frac{T^{in}}{T^{ex}}
        \end{array} \right)
\left( \begin{array}{c}  u_{i,B} \\ u_{i,A} \end{array} \right) .
\end{eqnarray}
Then, we can easily see that a solution can be chosen to satisfy either 
$u_{i,A} \equiv 0 $ or $u_{i,B} \equiv 0 $. 
Because $u_{1,B} = 0 $ to satisfy the boundary condition of edge states, 
we have a solution, 
\begin{eqnarray}
u_{i,A}= (-1)^{i-1} (~ \frac{T^{in}}{T^{ex}} ~)^{i-1} u_{1,A}.
\end{eqnarray}
Thus, in region of reciprocal lattice space where ${\bf k}$ satisfies 
$|T^{in}/T^{ex}| < 1 $, we have a dumping wave. 
The dumping factor $D({\bf k})$ is given by $T^{in}/T^{ex}$. 

Next, we have to show that a wave vector satisfying $|D({\bf k})|<1$
exists in a given network whose $T^{in}, T^{ex}$ and $u_{i,j}$ are a
scalar.  
But, we have assumed that a zero energy states exist on $\Lambda^{1}$,
which is given by {\bf k$_{0}$} satisfying $T^{in}({\bf k}_{0})=0$. 
Namely, this state is a localized eigen-state. 
Thus existence of a solution with $D=0$ is assumed. 
In usual networks with finite number of bonds for each site, 
$T^{in}$ and $T^{ex}$ are analytic functions of $k_j$ ($j=1,\cdots,N_0$). 
Hence, one can assume the continuity of $T^{in}$ (and $T^{ex}$) as a
function of $k_j$. 
Then we have solutions with $0<|D|<1$, which gives the degenerate edge
states.  

The graphene and the 3-5 network shown in this paper correspond to this
simplest case.  
In case of graphene, $T^{ex}=-1$ and 
$T^{in}= -2 \cos \frac{k}{2}$. 
Hence, the dumping factor $D$ becomes  $2\cos \frac{k}{2}$.
While, in case of the 3-5 network, 
$T^{ex}=-1$ and $T^{in}=-2 ( \cos \frac{k_{x}}{2} + \cos
\frac{k_{y}}{2}$).  
Hence, the dumping factor $D$ becomes 
$D = 2 ( \cos \frac{k_{x}}{2} + \cos \frac{k_{y}}{2}$).

So far, we have assumed that $T^{ex}, T^{in}$ and $u_{i,j}$ are scalars. 
In general, $T$s and $u_{i,j}$ could be a matrix and a vector,
respectively. 
Then there is a case which $T^{ex}$ become a singular matrix. 
Hence, it is impossible to derive analogue of Eqs.(\ref{t1}) and
(\ref{t2}) from Eq.(\ref{s1}).
But, we may be able to obtain analogous determination equations 
for an $N$-dimensional network. 

In case when $(T^{ex})^{-1}$ exist, 
we see that an eigenvector of each A-sites on $\Lambda_{i}$ 
is given by $D^{i-1}u_{1,A}$ and $u_{i,B} = 0$. 
Thus, the eigenvector of each A-sites on $\Lambda_{i}$ is represented
by eigenvector of each A-sites on $\Lambda_{i-1}$ and a dumping factor
$D$ as,  
\begin{eqnarray}
{\bf u}_{i+1,A} = - D~ {\bf u}_{i,A}. 
\label{wf}
\end{eqnarray}
Substituting Eq.(\ref{wf}) to Eq.(\ref{s1}), we have, 
\begin{eqnarray}
 T^{in} {\bf u}_{i,A} = - D~ T^{ex} {\bf u}_{i,A}.
\end{eqnarray}
The dumping factor, $D$, is determined by this equation.
We can judge whether edge states emerge at an edge of 
the $N$-dimensional network from it.
When $D = 0$, the edge state is completely localized at the surface. 
When $ 0 < |D| < 1 $, the edge state becomes a dumping wave.   
When $|D| = 1$, the state coincides with a bulk state. 
When $  1 < |D| $, the solution becomes unphysical.

We apply this formulation to the 3-3 network.
The equation becomes 
\begin{eqnarray}
\left( 
    \begin{array}{cc}
       - (1+ \exp (2i k_{x}) &
    - \exp (i(k_{x}+k_{y})) \\
       -1    &
       - (1+ \exp (2i k_{y}))
    \end{array} 
\right)
\left( \begin{array}{c} u_{1,A} \\ u_{2,A} \end{array} \right)
= -D
\left( \begin{array}{cc} 
           0  &  0  \\
          -1  &  0  
        \end{array} \right)
\left( \begin{array}{c} u_{1,A} \\ u_{2,A} \end{array} \right).
\end{eqnarray}
With this equation, we obtain 
a dumping factor for the 3-3 network. 
It is given by $D= 4 \cos k_{x} \cos k_{y} $. 

\begin{multicols}{2}[]

\end{multicols}

\end{document}